\documentstyle[12pt]{article}
\begin{document}
\begin{center}
{\LARGE \bf The Big Bang in $T^3$ Gowdy Cosmological Models }
\vskip1cm
{Hernando Quevedo}\\
\vskip5mm

Instituto de Ciencias Nucleares\\
Universidad Nacional Aut\'onoma de M\'exico \\
A. P. 70--543, M\'exico, D. F. 04510,  M\'exico. \\
\end{center}

\begin{abstract}
We establish a formal relationship between stationary axisymmetric spacetimes
and $T^3$ Gowdy cosmological models which allows us to derive several preliminary
results about the generation of exact cosmological solutions and 
their possible behavior near the initial singularity. In particular, we 
argue that it is possible to generate a Gowdy model from its values at the
singularity and that this could be used to construct cosmological solutions
with any desired spatial behavior at the Big Bang. 
\end{abstract}

\section {Introduction}

The Hawking-Penrose theorems \cite{hawel} prove that singularities are a 
generic characteristic of Einstein's equations. 
These theorems establish an equivalence between geodesic incompleteness and
the blow up of some curvature scalars (the singularity) and allow us to determine
the region (or regions) where singularities may exist. Nevertheless, these theorems
say nothing about the nature of the singularities. This question is of great
interest especially in the context of cosmological models, where the initial 
singularity (the ``Big Bang") characterizes the ``beginning" of the evolution 
of the universe. 

The first attempt to understand the nature of the Big Bang was made by Belinsky, 
Khalatnikov and Lifshitz \cite{bkl}. They argued that the generic Big Bang is 
characterized by the mixmaster dynamics of spatially homogeneous Bianchi cosmologies
of type VIII and IX. However, there exist counter-examples suggesting
that the mixmaster behavior is no longer valid in models with more than 
three dynamical degrees of freedom \cite{mike}. Several alternative behaviors 
have been suggested \cite{barrow}, but during the last two decades investigations
have concentrated on the so--called asymptotically velocity term dominated (AVTD)
behavior according to which, near the singularity, each point in space is
characterized by a different spatially homogeneous cosmology \cite{eardley}. 
The AVTD behavior of a given cosmological metric is obtained by solving
the set of ``truncated" Einstein's equations which are the result of neglecting 
all terms containing spatial derivatives and considering only the terms with
time derivatives. 

Spatially compact inhomogeneous spacetimes admitting two commuting spatial
Killing vector fields are known as Gowdy cosmological models \cite{gowdy}. Recently,
a great deal of attention has been paid to these solutions as favorable 
models for the study of the asymptotic behavior towards the initial cosmological 
singularity. Since Gowdy spacetimes provide the simplest inhomogeneous cosmologies,
it seems natural to use them to analyze the correctness of the AVTD behavior. In 
particular, it has been proved that all polarized Gowdy models belong to the class
of AVTD solutions and it has been conjectured that the general (unpolarized) models
are AVTD too \cite{mon1}. 

In this work, we focus on $T^3$ Gowdy cosmological models and 
present the Ernst representation of the corresponding field equations. We
show that this representation can be used to explore different types of solution
generating techniques which have been applied very intensively to generate 
stationary axisymmetric solutions. We use this analogy to apply several known
theorems to the case of Gowdy cosmological solutions. In particular, we 
use these results to show that all polarized $T^3$ Gowdy models preserve 
the AVTD behavior at the initial singularity and that ``almost" all 
unpolarized $T^3$ Gowdy models can be generated from a given polarized seed
solution by applying the solution generating techniques. We also analyze the 
possibility of generating a polarized model if we specify {\it a priori} any 
desired value of its Ernst potential at the initial singularity. We also argue
that this method could be used to generate a cosmological model starting from
its value at the Big Bang. 

\section{Gowdy $T^3$ Cosmological Models}

Gowdy cosmological models are characterized by the existence of two commuting
spatial Killing vector fields, say, $\eta_I = \partial/\partial\sigma$ and 
$\eta_{II} = \partial/\partial\delta$ which define a two parameter spacelike 
isommetry group. Here $\sigma$ and $\delta$ are spatial coordinates delimited
by $0 \leq \sigma, \ \delta \leq 2\pi$ as a consequence of the space topology.
In the case of a $T^3$-topology, the line element for unpolarized Gowdy models 
\cite{gowdy} can be written as
\begin{equation}
ds^2= e^{-\lambda/2} e^{\tau/2} ( - e^{-2\tau} d\tau^2 + d\theta^2)
+ e^{-\tau} [e^P (d\sigma + Q d\delta)^2 
+ e^{-P} d\delta^2] \ ,
\label{t3}
\end{equation}
where the functions $\lambda$,  $P$ and $Q$ depend on the coordinates 
$\tau$ and $\theta$ only, with $ \tau \geq 0$ and $0 \leq \theta \leq 2\pi$.
In the special case $Q=0$, 
the Killing vector fields $\eta_I$ and $\eta_{II}$ become hypersurface orthogonal to
each other and the metric (\ref{t3}) describes the polarized
$T^3$ Gowdy models. 

The corresponding Einstein's vacuum field equations consist
of a set of two second order differential equations for $P$ and
$Q$
\begin{equation}
P_{\tau\tau} - e^{-2\tau} P_{\theta\theta} - e^{2P}(Q_\tau^2 -
e^{-2\tau}Q_\theta^2) = 0 \ ,
\label{t3eqp}
\end{equation}

\begin{equation}
Q_{\tau\tau} - e^{-2\tau} Q_{\theta\theta} + 2(P_\tau Q_\tau -
e^{-2\tau}P_\theta Q_\theta) = 0 \ ,
\label{t3eqq}
\end{equation}
and two first order differential equations for $\lambda$ 
\begin{equation}
\lambda_\tau = P_\tau^2 + e^{-2\tau}P_\theta^2 + e^{2P} (Q_\tau^2 +
e^{-2\tau}Q_\theta^2) \ , 
\label{t3eqlam1}
\end{equation}

\begin{equation}
\lambda_\theta = 2(P_\theta P_\tau +
e^{2P}Q_\theta Q_\tau)  \ . 
\label{t3eqlam2}
\end{equation}
The set of equations for $\lambda$ can be solved by quadratures
once $P$ and $Q$ are known, because the integrability condition
$\lambda_{\tau\theta} = \lambda_{\theta\tau}$ turns out to be 
equivalent to Eqs.(\ref{t3eqp}) and (\ref{t3eqq}). 

To apply the solution generating techniques to the $T^3$ Gowdy models
it is useful to introduce the Ernst representation of the field 
equations. To this end, let us introduce a new coordinate  
$t=e^{-\tau}$ and a new function $R=R(t,\theta)$ 
by means of the equations
$ 
R_t = t e^{2P} Q_\theta  ,\ R_\theta = t e^{2P} Q_t
$. 
Then, the field equations (\ref{t3eqp}) and (\ref{t3eqq}) 
can be expressed as
\begin{equation}
t^2\left( P_{tt} + {1 \over t} P_t - P_{\theta\theta} \right) 
+ e^{-2P} (R_t^2 - R_\theta^2) = 0 \ ,
\label{eqpr}
\end{equation}

\begin{equation}
t e^P \left( R_{tt} + {1 \over t} R_t - R_{\theta\theta} \right)
- 2 [(t e^P)_t R_t - (te^P)_\theta R_\theta ] = 0  \ .
\label{eqqr}
\end{equation}
Furthermore, this last equation for $R$ turns out to be
identically satisfied if the integrability condition 
$R_{t\theta} = R_{\theta t}$ is fulfilled.
We can now introduce the complex Ernst potential $\epsilon$ and
the complex gradient operator $D$ as
\begin{equation}
\epsilon = t e^P + i R \ , \qquad {\rm and} \qquad
D= \left({\partial \over \partial t} \ , \ 
         i {\partial \over \partial \theta} \right) \ ,
\label{ernstpot}
\end{equation}
which allow us to write the main field equations in the 
{\it Ernst-like representation}
\begin{equation}
Re(\epsilon)\left(D^2\epsilon + {1\over t} D t\, D\epsilon \right)
 - (D\epsilon)^2 = 0 \ .
\label{ernstt3}
\end{equation}
It is easy to see that the field equations (\ref{eqpr}) and 
(\ref{eqqr}) 
can be obtained as the real and imaginary part of the Ernst
equation (\ref{ernstt3}), respectively. 

The particular importance of the Ernst representation (\ref{ernstt3}) is 
that it is very appropriate to investigate the symmetries of the 
field equations. In particular, the symmetries of the Ernst equation 
for stationary axisymmetric spacetimes have been used to develop the
modern solution generating techniques \cite{dh}. Similar studies can be carried 
out for any spacetime possessing two commuting Killing vector fields.
Consequently, it is possible to apply the known techniques (with some
small changes) to generate new solutions for Gowdy cosmological models.
This task will treated in a forthcoming work. Here, we will use the
analogies which exist in spacetimes with two commuting Killing vector 
fields in order to establish some general properties of Gowdy cosmological
models.

An interesting feature of Gowdy models is its behavior at the initial 
singularity which in the coordinates used here corresponds to the 
limiting case $\tau \rightarrow \infty$. The asymptotically velocity
term dominated  (AVTD) behavior has been conjectured as a characteristic 
of spatially inhomogeneous Gowdy models. This behavior implies that, at the
singularity, all spatial derivatives in the field equations can be neglected
in favor of the time derivatives. For the case under consideration, it
can be shown that the AVTD solution can be written as \cite{mon2}
\begin{equation} 
P = \ln[\alpha (e^{-\beta\tau} + \zeta^2 e^{\beta\tau})] \ , \qquad
Q= {\zeta \over \alpha ( e^{-2\beta\tau} + \zeta^2) } + \xi \ ,
\label{avtd}
\end{equation}
where $\alpha, \ \beta,\ \zeta$ and $\xi$ are arbitrary functions of $\theta$. At the
singularity, $\tau\rightarrow\infty$, the AVTD solution behaves as $P\rightarrow \beta
\tau$ and $Q\rightarrow Q_0 = 1/(\alpha\zeta) + \xi$. It has been shown \cite{ismon} that 
that all polarized ($Q=0$) $T^3$ Gowdy models have the AVTD behavior, while for 
unpolarized $(Q\neq 0$) models this has been conjectured. We will see in the following
section that these results can be confirmed by using the analogy with stationary axisymmetric
spacetimes.

\section{Analogies and general results}

Consider the line element for stationary axisymmetric spacetimes in the Lewis-Papapetrou 
form
\begin{equation}
ds^2 = - e^{2\psi} (dt + \omega d\phi)^2 + e^{-2\psi} [e^{2\gamma}(d\rho^2 + dz^2) 
+ \rho^2 d\phi^2]  \ , 
\label{star}
\end{equation}
where $\psi,\ \omega,$ and $\gamma$ are functions of the nonignorable coordinates 
$\rho$ and $z$. The ignorable coordinates $t$ and $\phi$ are associated with 
two Killing vector fields $\eta_I = \partial/\partial t$ and 
$\eta_{II} = \partial/\partial \phi$. The field equations take the form
\begin{equation}
\psi_{\rho \rho} + {1\over \rho}\psi_\rho + \psi_{zz} + 
{e^{4\psi}\over 2\rho^2}(\omega_\rho^2 + \omega_z^2) = 0 \ ,
\label{psi}
\end{equation}
\begin{equation}
\omega_{\rho \rho} - {1\over \rho}\omega_\rho + \omega_{zz} +
 4 (\omega_\rho\psi_\rho + \omega_z\psi_z ) = 0 \ ,
\label{omega}
\end{equation}
\begin{equation}
\gamma_\rho = \rho(\psi_\rho^2 - \psi_z^2) - 
{e^{4\psi}\over 4 \rho^2}(\omega_\rho^2 -  \omega_z^2)\ ,
\label{gammarho}
\end{equation}
\begin{equation}
\gamma_z = 2\rho \psi_\rho\psi_z - {1\over 2\rho} e^{4\psi} \omega_\rho\omega_z \ .
\label{gammaz}
\end{equation}
Consider now the following coordinate transformation $(\rho, t)\rightarrow (\tau,\sigma)$
and the complex change of coordinates $(\phi,z)\rightarrow (\delta,\theta)$ defined 
by
\begin{equation}
\rho= e^{-\tau}, \quad t=\sigma, \quad z=i\theta, \quad \phi = i\delta,
\label{trans1}
\end{equation}
and introduce the functions $P$, $Q$ and $\lambda$ by means of the relationships
\begin{equation}
\psi ={1\over 2}(P-\tau), \quad Q = i\omega, \quad \gamma = {1\over 2}
\left(P-{\lambda\over 2} -{\tau\over 2}\right)\ .
\label{trans2}
\end{equation}
Introducing Eqs.(\ref{trans1}) and (\ref{trans2}) into the line element (\ref{star}),
we obtain the Gowdy line element (\ref{t3}), up to an overall minus sign. Notice that
this method for obtaining the Gowdy line element from the stationary axisymmetric one
involves real as well as complex transformations at the level of coordinates and
metric functions. It is, therefore, necessary to demand that the resulting metric
functions be real. Indeed, one can verify that the action of the 
transformations (\ref{trans1}) and
(\ref{trans2}) on the field equations (\ref{psi})-(\ref{gammaz}) yields exactly 
the field equations (\ref{t3eqp})-(\ref{t3eqlam2}) for the Gowdy cosmological 
models. This is an interesting property that allows us to generalize several 
results known for stationary axisymmetric spacetimes to the case of Gowdy spacetimes.

The counterparts of static axisymmetric solutions ($\omega=0)$ are the polarized 
($Q=0)$ Gowdy models. For instance, the Kantowski-Sachs \cite{ks} cosmological model is the
counterpart of the Schwarzschild spacetime, one of the simplest static solutions.
Furthermore, it is well known that the field equations for static axisymmetric
spacetimes are linear and there exists a general solution which can be 
generated (by using properties of harmonic functions) from the Schwarzschild one 
\cite{queprl}. According to the analogy described above, this implies the following

\noindent
{\bf Lemma 1}: All polarized $T^3$ Gowdy cosmological models can be generated from 
the Kantowski-Sachs solution.

The method for generating polarized Gowdy models can be briefly explained in 
the following way. If we introduce the time  coordinate $t = e ^{-t}$,
 Eq. (\ref{t3eqp}) becomes $\Delta P = 0$ with
$\Delta = \partial_{tt} + t^{-1}  \partial_{t} - \partial_{\theta\theta}$. 
It can easily be verified that the operators $L_{1} = t^{-1} \partial_{t} (t\partial_{t})$ 
and $L_{2} = \partial_{\theta}$ commute with the operator $\Delta$. 
Then if a solution $P_{0}$ is known $(\Delta 
P_{0} = 0)$, the action of the operator $L_{i}\ (i=1,2)$ on $P_{0}$ generates  new solutions
 $\tilde P_i$, i.e., if $ \tilde P_{i}= L_{i} P_{0}$ 
then $\Delta \tilde P_{i} = \Delta L_{i} P_{0} = 0$. 
This procedure can be repeated as many times as desired, generating an infinite number 
of solutions whose sum
with arbitrary constant coefficients represents the general solution. 
An important property of the action of $L_{i}$ on a given solution $P_{0}$ is that 
it preserves the behavior of $P_{0}$ for
$ \tau\rightarrow \infty$. If we choose the Kantowski-Sachs spacetime as the 
seed solution $P_{0}$, the general solution will be AVTD. Hence, 
as a consequence of Lemma 1, we obtain 

\noindent
{\bf Lemma 2}: All polarized $T^{3}$ Gowdy cosmological models are AVTD.

We now turn to the general unpolarized $(Q \neq 0)$ case. The solution generating 
techniques have been applied  intensively to generate stationary solutions from static ones. 
In particular, it has been shown that ``almost" all stationary axisymmetric solutions can 
be generated from a given static solution (the Schwarzschild spacetime, 
for instance) \cite{dh}. The term ``almost" means that there exist ``critical" points 
where the field equations are not well defined and, therefore, the solution  generating techniques can not bee applied. Using the analogy with Gowdy models, we obtain

\noindent
{\bf Lemma 3:} ``Almost" all unpolarized $T^{3}$ Gowdy cosmological solutions can be 
generated from a given polarized seed solution.

As in the previous case, it can be shown that the solution generating techniques 
preserve the asymptotic behavior of the seed solution. 
If we take the Kantowski-Sachs spacetime as seed solution and apply Lemma 3, we obtain

\noindent
{\bf Lemma 4:} ``Almost" all $T^{3}$ unpolarized Gowdy models can be generated from the Kantowski-Sachs solution and are AVTD.

It should be mentioned that all the results presented in Lemma 1 - 4 must be treated as ``preliminary". To ``prove" them we have used only the analogy between 
stationary axisymmetric spacetimes and Gowdy cosmological models, 
based on the transformations (\ref{trans1}) and (\ref{trans2}).
A rigorous proof requires a more detailed investigation and analysis of the symmetries 
of the equations (\ref{t3eqp})-(\ref{t3eqlam2}), especially in their Ernst 
representation (\ref{ernstt3}).

\section{The Big Bang}

The behavior of the Gowdy $T^{3}$ cosmological models at the initial singularity 
$(\tau\rightarrow \infty)$ is
dictated by the AVTD solution (\ref{avtd}). 
Although this is not the only possible case for the Big Bang, there are physical reasons to believe that this is a  suitable scenario for the simplest case of inhomogeneous cosmological models. But an inhomogeneous Big Bang has two different aspects. The first one
 concerns the temporal behavior as the singularity is approached. The second aspect is 
related to the spatial inhomogeneities which should be present during the Big Bang. 
If we accept the AVTD behavior, the first aspect of the problem becomes solved by the 
AVTD solution (\ref{avtd}), which determines the time dependence at the Big Bang. 
However, the spatial dependence remains undetermined as it is given by the arbitrary 
functions $\alpha,\ \beta,\ \zeta$ and $\xi$ which can be specified only once a solution is known.  The question arises whether it is possible to have a solution with 
any desired spatial behavior at the Big Bang. We will see that the answer to this 
question is affirmative.

Using the solution generating techniques for stationary axisymmetric spacetimes it has been 
shown that any solution can be generated from its values on the axis of symmetry $(\rho = 0)$
\cite{dh}. Specific procedures have been developed that allow us to construct any 
solution once the value of the corresponding Ernst potential is given at the axis
\cite{sibg}. 
 On the other hand, the analogy with Gowdy models determined by Eqs.(\ref{trans1}) 
and (\ref{trans2}) indicates that the limit $\rho \rightarrow 0$ is equivalent to 
$\tau \rightarrow\infty$. To see that this equivalence is also valid for specific 
solutions we write the counterpart of the AVTD solution 
(\ref{avtd}) in the coordinates $\rho$ and $z$, according to the Eqs.(\ref{trans1}) 
and (\ref{trans2}). Then, we obtain
\begin{equation}
\psi = {1\over 2}\ln[ \alpha (\rho^{1+\beta} + \zeta^2\rho^{1-\beta})] , 
\qquad \omega = {\zeta\over \alpha(\rho^{2\beta} + \zeta^2)} + \xi\ ,
\label{avtdrho}
\end{equation}
where we have chosen the arbitrary functions $\zeta$ and $\xi$ such that $\omega$
becomes a real function. One can verify that the solution (\ref{avtdrho}) satisfies the 
corresponding ``AVTD equations" (\ref{psi}) and (\ref{omega}) (dropping the 
derivatives with respect to $z$) for the stationary case. Consequently, the behavior of 
stationary axisymmetric solutions at the axis $(\rho \rightarrow 0)$ is equivalent to the 
behavior at $T^{3}$ Gowdy models at the Big Bang.

Thus, if we consider the asymptotic behavior of the Ernst potential (\ref{ernstt3})
 for the AVTD solution (\ref{avtd})
\begin{equation}
\epsilon(\tau\rightarrow \infty) \rightarrow \epsilon_0 = 
e^{\tau(\beta -1)}[ 1 + Q_0 e^{\tau(\beta -1)}]\ , \qquad Q_0 = {1\over \alpha\zeta} 
+ \xi \ , 
\label{e0}
\end{equation} 
and specify $\alpha,\ \beta,\ \zeta$ and $\xi$ as functions of the spatial  
coordinate $\theta$, it is possible to derive the corresponding unpolarized 
$(Q \neq 0)$ solution by using the solution generating techniques. In other words, we can 
construct a Gowdy model with any desired behavior at the Big Bang. 
This is an interesting possibility which could be used to analyze physically reasonable 
scenarios for the  Big Bang.

Of course, a more detailed study of the solution generating techniques for Gowdy 
cosmological models is necessary in order to provide a rigorous proof of these
results and to construct realistic models for the 
Big Bang. Here we have used the analogy between stationary axisymmetric spacetimes 
and $T^{3}$ Gowdy cosmological models to show that this is a task that in principle 
can be solved.

\section{Acknowledgements}

It is a pleasure to dedicate this work to Prof. H. Dehnen and Prof. D. Kramer on their
60-th birthday. I would like to thank O. Obreg\'on and M. Ryan for an ongoing collaboration
that contributed to this work. Also, support by DLR-CONACyT, and 
DGAPA-UNAM, grant 121298, is gratefully acknowledged.

\end{document}